\documentclass[twoside]{article}
\usepackage{fleqn,espcrc2}

\usepackage{graphicx}
\usepackage[figuresright]{rotating}

\newcommand{\AmS}{{\protect\the\textfont2
  A\kern-.1667em\lower.5ex\hbox{M}\kern-.125emS}}

\title{Theory of vortex lattice effects on STM spectra
in d-wave superconductors}

\author{Alexander S.Mel'nikov
\address{Institute for Physics of the Microstructures,
Russian Academy of Sciences,\\
 GSP-105,  Nizhny Novgorod, 603600, Russia}%
\thanks{This work was supported, in part, by the Russian Foundation for
Fundamental Research (Grant Nos. 99-02-16188 and 97-02-17437).}}

\begin{document}
\pagestyle{empty}

\begin{abstract}
Theory of scanning tunneling spectroscopy of low energy quasiparticle
(QP) states in vortex lattices of d-wave superconductors is developed
taking account of the effects caused by an extremely large extension
of QP wavefunctions in the nodal directions and the band structure in the
QP spectrum.  The oscillatory structures in STM spectra, which correspond
to van Hove singularities are analysed.  Theoretical calculations carried
out for finite temperatures and scattering rates are compared with recent
experimental data for high-$T_c$  cuprates.  \vspace{1pc}
\end{abstract}
\maketitle
The electronic structure of the mixed state in d-wave superconductors
reveals a number of fundamentally new features
(see \cite{volovik,me,ander,janko,franz,kita}
and references therein) as
compared to the case of s-wave
compounds, where low lying quasiparticle (QP) states are bound to the
vortex core and are weekly perturbed by the presence of neibouring
vortices at magnetic fields ${H\ll H_{c2}}$.  The vanishing
pair potential in the nodal directions results in the extremely
large extension of QP wavefunctions which are sensitive to
the superfluid velocity (${\bf V}_s$) fields of all vortices and, thus,
the electronic structure is influenced by the vortex lattice geometry.
The resulting peculiarities of the local density of states (DOS) can be
detected, e.g., by a scanning tunneling microscope (STM).  In this paper
we focus on the theory of scanning tunneling spectroscopy of low energy
QP states in vortex lattices of d-wave superconductors and compare the
theoretical calculations with recent experimental data \cite{YBCO,BSCCO}
for high-$T_c$ cuprates, where the dominating order parameter is
believed to be of d-wave symmetry.  Hereafter we assume the Fermi surface
(FS) to be two-dimensional (2D),
take the gap function in the form
${\Delta_{\bf k}=2\Delta_0 k_xk_y/k_F^2}$
(the $x$ axis makes an angle $\pi/4$ with the $a$ axis of
the $CuO_2$ planes). Let us orient ${\bf H}$ along the $\it c$
axis (${H_{c1}\ll H\ll H_{c2}}$)
and consider two types of vortex lattices:  (I) rectangular
lattice with primitive translations ${\bf a}_1=a{\bf x}_0$, ${\bf
a}_2=\sigma a{\bf y}_0$; (II) centered rectangular lattice  with ${\bf
a}_1=a{\bf x}_0$, ${\bf a}_2=a ({\bf x}_0/2-\sigma {\bf y}_0)$, where
$H\sigma a^2=\phi_0$ is the flux quantum,
and ${\bf x}_0$, ${\bf y}_0$, ${\bf z}_0$
are the unit vectors of the coordinate system.

\begin{figure}[b]
\leavevmode
\includegraphics[width=0.7\linewidth]{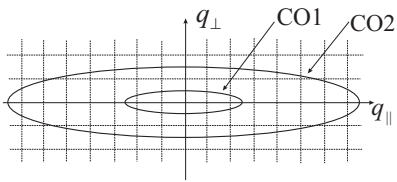}
\caption{Cyclotron orbits (CO1, CO2) and MBZ
boundaries for a square lattice.
$(q_{\parallel},q_{\perp})$ defines a
coordinate system whose origin is at the node, with $q_{\perp}$
($q_{\parallel}$) normal (tangential) to the FS.
}
\label{fig1}
\end{figure}

\begin{figure}[b]
\leavevmode
\includegraphics[width=1.0\linewidth]{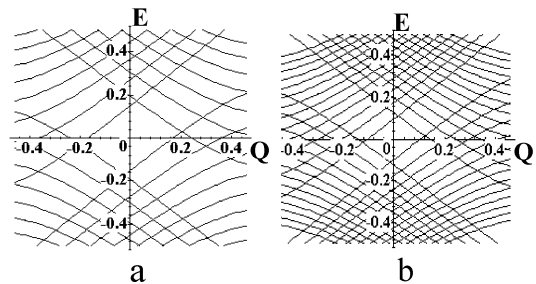}
\caption{Energy branches for
$\pi\sigma \alpha=50$~(a) and
$\pi\sigma \alpha=100$~(b)}
\label{fig2}
\end{figure}

{\it Van Hove singularities.}
Our consideration is based on the analysis of the Bogolubov-de Gennes
(BdG) equations for low energy excitations with momenta close
to a certain gap node direction
(e.g., ${\bf k}_1=k_F{\bf x}_0$):
${(\hat H_0+\hat H^\prime)\hat g=\varepsilon\hat g}$,
where  ${\hat g=(u,v})$ is the QP wavefunction,
${\hat H_0=V_F\hat\sigma_z\hat p_x+V_\Delta\hat\sigma_x\hat p_y}$,
$\hat\sigma_x,\hat\sigma_z$ are the Pauli matrices,
${\hat H^\prime=MV_FV_{sx}(1+\hat\sigma_z)+MV_\Delta V_{sy}\hat\sigma_x}$,
$M$ is the electron effective mass,
${\hat {\bf p}=-i\hbar\nabla-e{\bf A}/c}$,
${V_F=\hbar k_F/M}$, ${V_\Delta=2\Delta_0/(\hbar k_F)}$,
${{\bf H}=-H{\bf z}_0}$,
${{\bf A}=Hy{\bf x}_0}$,
${\bf V}_s=(V_{sx},V_{sy})$.
The spectrum of the Dirac Hamiltonian $\hat H_0$
can be obtained using the usual quantization rule
for a cyclotron orbit (CO) area \cite{ander,janko}.
The periodic potential $\hat H^\prime$ removes
the degeneracy of the discrete energy levels
with respect to the CO center and induces
a band structure in the spectrum \cite{me,franz,kita}.
The general solution can be written in the form
of a magnetic Bloch wave:

\begin{equation}
\label{mbw}
\hat g=\sum\limits_n e^{ix(q_x+2\pi n/a)+2in\sigma q_y a}
\hat G(y-2n\sigma a, {\bf q}),
\end{equation}

\noindent where $n$ is an integer, and ${\bf q}$
is the quasimomentum lying within the
first magnetic Brillouin zone (MBZ): ${-\pi/(2a)<q_x<\pi/(2a)}$,
${-\pi/(2\sigma a)<q_y<\pi/(2\sigma a)}$.
The wavefunction $\hat G(y,{\bf q})$ is localized in the domain
with the size $L$ determined by ${\bf q}$ and energy values.
The potential $\hat H^\prime$ results in the splitting of the
CO near MBZ boundaries (see Fig.~\ref{fig1}) and
the spectrum consists of branches which correspond to the
splitted portions of the CO.
For large Dirac cone anisotropy ${\alpha=V_F/V_\Delta\gg 1}$
($\alpha=k_F\xi_0/2$) and ${\varepsilon< 0.5\varepsilon^*}$
(${\varepsilon^*=\pi\hbar V_F/a\sim \Delta_0\sqrt{H/H_{c2}}}$)
the harmonics in Eq.~(\ref{mbw}) do not overlap
($L<2\sigma a$) and one can replace
$\hat H^\prime$ by the potential $\langle\hat H^\prime\rangle_x$
averaged in the $x$ direction (see \cite{me}). Such a
simplification is a natural consequence of a small
size of the cyclotron orbit (CO1 in Fig.~\ref{fig1})
in the nodal direction as compared to the size of the MBZ.
The energy spectrum consists of branches
${\varepsilon_n(q_x=\pi Q/a)=\varepsilon^*E_n(Q,\sigma \alpha)}$,
which are displayed in Fig.~\ref{fig2}
in the first MBZ for
$\pi\sigma \alpha=50$ and $\pi\sigma \alpha=100$.
The number of energy branches
which cross the Fermi level can be determined as follows:
${N\sim 2q_{\parallel}^*/\delta q_{\parallel}
\sim 2\sqrt{\pi\sigma \alpha}}$,
where $q_{\parallel}^*$ is the minimum possible size of the CO
in the $q_{\parallel}$  direction and $\delta q_{\parallel}$ is
the distance between MBZ boundaries.
Each energy branch has an extremum as a function of the
momentum $q_x$ near the MBZ boundary at a certain
$\tilde\varepsilon_n$ (we neglect here additional extrema which
appear due to the exponentially small splitting of energy levels near
the points of intersection of the branches in the $E-Q$ plane).
Due to the one-dimensional (1D) nature of the low energy spectrum
the divergent contributions to the DOS take the form:
$\delta N(\varepsilon)\sim |\varepsilon-\tilde\varepsilon_n|^{-1/2}$
($\varepsilon>\tilde\varepsilon_n$ for energy minima and
$\varepsilon<\tilde\varepsilon_n$ for maxima).
The distance between these peaks
${\delta \varepsilon \sim \varepsilon^*/(2\sigma \alpha)}$
coincides with a characteristic energy scale
corresponding to van Hove singularities which occur
when the CO intersects MBZ boundaries in the $q_y$
direction (see Fig.~\ref{fig1}).
The crossover between 1D and 2D regimes in the band spectrum
occurs at $\varepsilon_c\sim 0.5\varepsilon^*$,
when the CO size in the $q_\perp$ direction becomes larger than
the size of the first MBZ (CO2 in Fig.~\ref{fig1}).
For $\varepsilon\stackrel{_>}{_\sim}\varepsilon_c$ the $q_y$-dependence
of energy becomes essential and results in the appearance of
2D critical points, i.e. 2D local maxima (or minima) and saddle-points.
Thus, instead of square-root van Hove singularities
we obtain a set of discontinuities and logarithmic peculiarities
($\delta N(\varepsilon)\sim -ln|1-\varepsilon/\tilde\varepsilon_n|$)
in the DOS, respectively.
Obviously, these 2D singularities are more sensitive to temperature
and finite lifetime effects and, consequently, the suppression of
the corresponding oscillatory structure in the
DOS should be stronger in the high energy regime.
The above analysis can be generalized for gap nodes
at  ${\bf k}=\pm k_F{\bf y}_0$: the corresponding energy scales take the
form ${\delta \varepsilon \sim 0.5\varepsilon^*/\alpha}$,
${\varepsilon_{c} \sim 0.5\varepsilon^*/\sigma}$.

Even in the low energy regime the DOS oscillations with the energy scale
$\delta \varepsilon$ are surely smeared due to a finite scattering rate
$\Gamma$ and temperature and can be observed only for a moderate
Dirac cone anisotropy and rather large magnetic fields.
Comparing our results with a numerical solution \cite{kita}
of the BdG equations for $\sigma=1$ and $\alpha=5/2$
we find that the above mechanism gives a good estimate of the energy
scale of the double-peak structure in the tunneling conductance at the
core center at $H/H_{c2}=0.3$ ($\delta\varepsilon\sim 0.1\Delta_0$) and
can explain the absence of this structure at low fields $H/H_{c2}=0.05$
due to temperature broadening
($T=0.1T_c>\delta\varepsilon\sim 0.05\Delta_0$).
In principle, van Hove singularities may account for
peaks with a large energy gap
${\sim\Delta_0/4}$ observed experimentally at the vortex
centers in YBaCuO \cite{YBCO} at $H\simeq 6T$ provided we assume
$\alpha\sim 1$.
Unfortunately the latter assumption is not consistent with
the results of thermal conductivity measurements \cite{thermo}
($\alpha\sim 14$), and, thus, the nature of the experimentally observed
peaks is still unclear.
It is also necessary to stress here that the critical points in the DOS
are a direct consequence of perfect periodicity and the introduction of
rather strong disorder surely remove these singularities.

{\it Zero-bias conductance.} Hereafter we neglect the DOS
oscillations, discussed above, and consider the
peculiarities of the zero-bias tunneling conductance $g({\bf r})$
starting from a modified semiclassical model proposed in \cite{me}.
According to this approach,
the Doppler shift of the QP energy, which plays important role for
$\varepsilon\stackrel{_<}{_\sim}\Delta_0\sqrt{H/H_{c2}}$, appears to be
averaged in the nodal direction due to an extremely large size of a
semiclassical wave packet in this energy interval.
Within such an approximation a diagonal (retarded) Green's function can
be written in the form:

\begin{equation}
\label{green}
G^R({\bf k},\varepsilon,{\bf r})=
\frac{\varepsilon+i\Gamma-\hbar{\bf k}_F {\bf V}_{av}+
\epsilon_{\bf k}}
{(\varepsilon+i\Gamma-\hbar{\bf k}_F {\bf V}_{av})^2-
\Delta_{\bf k}^2-\epsilon_{\bf k}^2},
\end{equation}

\noindent where $\epsilon_{\bf k}$ is the normal state electron
dispersion,
${\bf V}_{av}=\langle{\bf V}_s\rangle_x+\langle{\bf V}_s\rangle_y$.
The scattering rate $\Gamma$ should be
determined self-consistently:
${\Gamma=N(\Gamma,\varepsilon)/(2N_F\tau)}$
(Born limit),
${\Gamma=N_F\Gamma_u/N(\Gamma,\varepsilon)}$
(unitary limit),
where $2\tau$ and $N_F$  are the relaxation time
and DOS at the Fermi level in the normal state,
$\Gamma_u=n_{imp}/(\pi N_F)$,
$n_{imp}$ is the concentration of of point potential scatterers,
and ${N(\Gamma,\varepsilon)=-Im\int G^Rd^2k/(2\pi^3)}$
is the local DOS.
\begin{figure}[b]
\leavevmode
\includegraphics[width=0.8\linewidth]{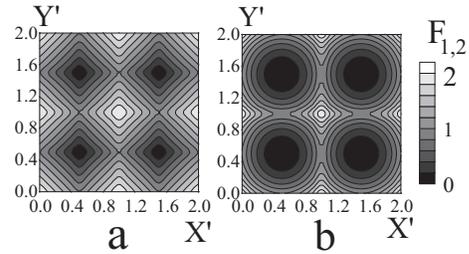}
\caption{
Contour plots of the functions $F_1$ (a) and $F_2$ (b)
which determine the spatial variation of the
zero-bias tunneling conductance for a square lattice
($x^{\prime}=x/a$, $y^{\prime}=y/a$).
}
\label{fig3}
\end{figure}
Let us first consider the effect of finite temperature on
the zero-bias conductance in the clean limit
($\Gamma\rightarrow 0$). The expression for the normalized conductance
reads:

$$
\tilde g({\bf r})=\frac{g({\bf r})}{g_N}=
\int\limits_{-\infty}^{+\infty}
\frac{N(\Gamma=0,\varepsilon)d\varepsilon}
{4N_FTcosh^2\left(\frac{\varepsilon}{2T}\right)}=\qquad\qquad
$$
\begin{equation}
\label{sigma}
\frac{T{\it ln}2}{\Delta_0}+
\frac{T}{2\Delta_0}\sum\limits_{i=x,y}{\it ln } cosh
\frac{\varepsilon_i^*\Phi_i}{4T}
\end{equation}

\noindent Here $g_N$ is the normal state conductance,
${\Phi_x=\Phi(x/R_x)}$, ${\Phi_y=\Phi(y/R_y)}$,
${\Phi(z)=2z-(2m+1)}$ for ${m<z<m+1}$ ($m$ is an integer),
$R_y$ ($R_x$) is the distance between the
lines parallel to the $x$ ($y$) axis and passing through the
vortex centers,
${\varepsilon_x^*=\pi\hbar V_F H R_x/\phi_0}$,
${\varepsilon_y^*=\pi\hbar V_F\sigma/R_y}$,
For type I (II) lattices we have
$R_x=a$, $R_y=\sigma a$ ($R_x=a/2$, $R_y=\sigma a$).
One can separate two qualitatively different regimes
in the behavior of the conductance:\\
(i) superflow dominated regime $T\ll \varepsilon^*_{x,y}$,

\begin{equation}
\label{sflow}
\tilde g
\simeq\frac{1}{8}\sqrt{\frac{\pi\sigma H}{2H_{c2}}}F_1(x,y),
\end{equation}

\noindent (ii) temperature dominated regime  $T\gg \varepsilon^*_{x,y}$,

\begin{equation}
\label{tempt}
\tilde g\simeq\frac{T{\it ln}2}{\Delta_0}+
\frac{\pi\Delta_0\sigma H}{32TH_{c2}}F_2(x,y),
\end{equation}

\noindent where $F_1(x,y)=|\Phi_x|(R_x/R_y)+|\Phi_y|$ and
$F_2(x,y)=\Phi_x^2(R_x/R_y)^2+\Phi_y^2$.
In Fig.~\ref{fig3} we display the contour plots of the
functions $F_1(x,y)$, $F_2(x,y)$ for a square lattice of type I
(which is close to the one observed experimentally
in YBaCuO \cite{YBCO}).
There are two consequences of an increase in temperature:
(i)~first, the spatial dimensions of peaks in the local DOS become rather
small comparing to the intervortex distance only for
${T >T^*\sim\Delta_0\sqrt{H/H_{c2}}}$;
(ii)~second, the amplitude of the peaks
appears to be essentially suppressed in the limit $T\gg T^*$.
For magnetic fields $H\sim 6T$ (which is typically
the field of STM experiment \cite{YBCO,BSCCO})
one obtains $T^*\sim 20K$.
Thus, we conclude that the finite
temperature effects can not explain neither the narrow zero-bias
conductance peaks observed in YBaCuO nor the absence of these peaks in
BiSrCaCuO at $T=4.2K$. To explain these experimental
facts it is necessary to take account of the finite lifetime
effects which can stronly influence on the behavior of the
DOS, as it follows from the results of
Refs.~\cite{barash,hirsch,frtec} obtained
on the basis of the usual semiclassical approach with a local Doppler
shift.  Starting from the modified semiclassical model (\ref{green})
we obtain the following expression for the tunneling conductance at $T=0$:

\begin{equation}
\tilde g= \frac{N(\Gamma,\varepsilon=0)}{N_F}=
\frac{\Gamma}{4\pi\Delta_0}\left(
4{\it ln}\frac{\Delta_0}{\Gamma}
+\sum\limits_{i=x,y}f_i\right)
\end{equation}
$$
f_i=\frac{\varepsilon^*_i|\Phi_i|}{\Gamma}
tan^{-1}\frac{\varepsilon^*_i|\Phi_i|}{2\Gamma}
-{\it ln}\left(1+\frac{(\varepsilon^*_i\Phi_i)^2}{4\Gamma^2}\right)
\qquad
$$

\noindent Obviously Born scatterers result only in a moderate change
of the DOS (see \cite{barash}) since the corresponding $\Gamma$ value
for $\Delta_0\tau\gg 1$ is very small comparing to
$\varepsilon^*_{x,y}$ and the conductance is given by Eq.~(\ref{sflow}).
On the contrary, in the unitary limit the expression (\ref{sflow}) is
valid only in the clean case ${\Gamma_u\ll \Gamma_{x,y}^*\sim
0.1\varepsilon_{x,y}^2/\Delta_0 }$ (for a square lattice
${\Gamma_{x,y}^*\sim 0.1\Delta_0 H/H_{c2}}$).  In the dirty limit
${\Gamma_u\gg \Gamma_{x,y}^*}$ we obtain:

\begin{equation}
\label{dirty}
\tilde g\simeq \tilde g (H=0)
\left(1+\frac{\Delta_0 H\sigma}{64\Gamma_u H_{c2}}F_2(x,y)\right),
\end{equation}

\noindent where $\tilde g (H=0)\simeq 0.5\sqrt{\Gamma_u/\Delta_0}$.
In the vicinity of each vortex center the local DOS exhibits a
fourfold symmetry with maxima along the nodal directions in a good
agreement with numerical calculations based on the Eilenberger theory
\cite{ichi}.
For $H=6T$ finite lifetime effects become substantial
if we assume $\Gamma_u\stackrel{_>}{_\sim}10^{-2}\Delta_0$.
Thus, our approach allows to explain
rather narrow conductance peaks (see Fig.~\ref{fig3}b)
observed near vortex centers in YBaCuO \cite{YBCO},
even without taking account of
the nontrivial structure of the tunneling matrix element,
discussed in \cite{frtec}.
With a further increase of the $\Gamma_u$ value the
amplitude of the peaks at the vortex centers vanishes:
${\delta \tilde g\sim \sqrt{\Delta_0/\Gamma_u}(H/H_{c2})}$.
Such a high sensitivity of the $\delta\tilde g$ value to finite lifetime
effects can probably explain the difficulties in the observation
of these peaks in the mixed state of BiSrCaCuO \cite{BSCCO}.
Note in conclusion that according to Eq.~(\ref{dirty}) the spatially
averaged DOS in the dirty limit varies as $H$ rather than $HlnH$ (the
latter dependence has been predicted in \cite{barash,hirsch} within the
semiclassical approach taking account of the local ${\bf V}_s$
value).

I am pleased to acknowledge useful discussions with
Dr.N.B.Kopnin, Dr.Yu.S.Barash,
Dr.A.A.Andronov, Dr.I.D.Tokman, and Dr.D.A.Ryndyk.

\end{document}